# Superconductivity at 27 K in tetragonal FeSe under high pressure

Yoshikazu Mizuguchi[1,2], Fumiaki Tomioka[1], Shunsuke Tsuda[1,3],
Takahide Yamaguchi[1] and Yoshihiko Takano[1,2]
1.*National Institute for Materials Science, 1-2-1, Sengen, Tsukuba, 305-0047, Japan*
2.*University of Tsukuba, 1-1-1, Tennodai, Tsukuba, 305-8577, Japan*
3.*WPI-MANA-NIMS, 1-1, Namiki, Tsukuba, 305-0044, Japan*

Abstract

A huge enhancement of the superconducting transition temperature $T_c$ was observed in a tetragonal FeSe superconductor under high pressure. The onset temperature was as high as 27 K at 1.48 GPa and the pressure coefficient showed an extremely high value of 9.1 K/GPa. The upper critical field $H_{c2}$ was estimated to be ~ 72 T at 1.48 GPa. Because of the high $H_{c2}$, the FeSe system can be applied to superconducting wire rods.

Since the great discovery of cuprate high-$T_c$ superconductors, over 20 years have passed. Recently many researchers have made much effort focused on non-cuprate, new superconductors such as $MgB_2$ with a $T_c$ of 39 K.[1,2,3]

The discovery of a new superconductor $LaFeAsO_{1-x}F_x$ with a $T_c$ of 26 K[4] was a great surprise. Soon after the discovery, active experimental and theoretical studies on this system followed, and various iron-based superconductors, which contain FeAs layers, have been reported.

The first series of iron-based superconductors is the LnFeAsO (Ln = Lanthanides) system with a tetragonal ZrCuSiAs structure. The superconducting transitions were obtained by either F doping to the O site or the deficiency of O ion in this system. The $T_c$ of $LaFeAsO_{1-x}F_x$ is 43 K under high pressure[5]. By the substitution of smaller Lanthanide ions such as Nd and Sm for La ions, $T_c$ increased to around 55 K due to chemical pressure[6,7]. The second series is $AFe_2As_2$ (A = Ba, Sr, Ca, Eu) system with a tetragonal $ThCr_2Si_2$ structure. $Ba_{1-x}K_xFe_2As_2$ exhibited the highest $T_c$ of 38 K[8]. The third series is LiFeAs system with a $T_c$ of 18 K[9,10]. The location of the Li ions in LiFeAs is similar to that of half of the Fe ions in $Fe_2As$ with $Cu_2Sb$ structure.

Very recently, a new type of iron-based superconductor, tetragonal FeSe with a $T_c$ of 8 K, was reported[11]. The crystal structure of tetragonal FeSe is the simplest in iron-based superconductors. We expect that the detailed investigation of FeSe system will provide important clues to understand the superconducting mechanisms on iron-based superconductors. Here, we report the superconducting properties of tetragonal FeSe, and dramatic enhancement of $T_c$ and upper critical field $H_{c2}$ by applying pressure.

Polycrystalline samples of FeSe were prepared using the solid state reaction method. High purity powders of Fe (99.9% up) and Se (99.999%) were mixed with nominal compositions and sealed into evacuated quartz tubes. Then the powders were heated at 680 °C for 12 hours. The obtained powders were reground and pressed into pellets. The pellets were heated at 680 °C for 12 ~ 30 hours.

We characterized the obtained samples by powder X-ray diffraction using CuK$_\alpha$ radiation. Temperature dependence of magnetization was measured using SQUID magnetometer at $H$ = 10 Oe. Resistivity measurements were performed using the four terminal method from room temperature to 2 K. Gold wires of 25 μm in diameter were attached to the sample using silver paste. Hydrostatic pressures were generated by a BeCu/NiCrAl clamped piston-cylinder cell. The sample was immersed in a fluid pressure transmitting medium of Fluorinert (FC70:FC77 = 1:1) in a Teflon cell. The pressure at low temperature was estimated from the superconducting transition temperature of Pb.

The X-ray diffraction pattern is shown in Fig. 1. All the peaks were well indexed using $P4/nmm$ space group. The calculated lattice constants were $a$ = 3.7696(6) Å and $c$ = 5.520(1) Å. These values were almost the same as those reported by G. Hägg and A. L. Kindstrom in 1933[12]. The small peaks of hexagonal FeSe (NiAs-type) were observed as an impurity phase.

Figure 2 shows the temperature dependence of magnetization of zero-filed cooling (ZFC) and field cooling (FC) modes. Below $T_C^{mag}$ ~ 8 K, we found clear diamagnetic responses corresponding to superconductivity. The shielding fraction was estimated to be about 11 %.

Figure 3 shows the temperature dependence of resistivity from room temperature to 2 K under pressure. We found two slight humps in the resistivity around 50 K and 200 K at 0.00 GPa, which may correspond to the structural phase transition and the magnetic transition, respectively, as reported for FeAs compounds. By applying pressure, the resistivity at 300 K decreases systematically and the humps become smaller. External pressure may suppress the structural transition and the magnetic transition.

Figure 4 shows magnified temperature dependence of resistivity at low temperatures. The superconducting transition temperatures are plotted as a function of applied pressure in Fig. 5. At 0.00 GPa, the resistivity starts to decrease due to superconductivity at $T_c^{onset}$ = 13.5 K, and drops to zero at $T_c^{zero}$ = 7.5 K. The zero resistivity temperature $T_c^{zero}$ = 7.5 K almost corresponds to the $T_c^{mag}$ = 8 K obtained in magnetization measurement. $T_c^{onset}$ and $T_c^{zero}$ increase drastically with increasing pressure. Surprisingly, $T_c^{onset}$ and $T_c^{zero}$ were enhanced to 27 K and 13.5 K under a pressure of 1.48 GPa, respectively. The pressure coefficient of FeSe was estimated to be 9.1 K/GPa from the onset temperature. This value is the largest in reported iron-based superconductors[5,13]. So far, we have not yet observed the optimal $T_c$ up to 1.48 GPa. It would be possible to obtain a higher $T_c$ by applying further pressure. Moreover, chemical pressure may also achieve a higher $T_c$ in the FeSe system.

We note a very interesting behavior in Fig. 4. The width of superconducting transition $\Delta T_c$ at 0.42 GPa is smaller than that of 0.00 GPa. In general, superconducting transition becomes broader with increasing pressure, but the transition becomes sharper with increasing pressure in FeSe. Pressure higher than 0.99 GPa makes the transition broader. We assume that the pressure would clear off something interfering with realizing ideal superconductivity, such as structural phase transition or magnetic transition. The optimal pressure thus may make FeSe layers ideal for superconductivity. To investigate the origin of this behavior, more experimental and theoretical studies are required.

Figure 6 (a) and (b) show the temperature dependence of resistivity under high pressures and magnetic fields up to 7 T with an increment of 1 T. With increasing magnetic field, superconducting transitions shift to lower temperatures and are slightly broadened. To estimate $H_{c2}$ and irreversible field $H_{irr}$, $T_c^{onset}$, $T_c^{zero}$ and the midpoint temperature $T_c^{mid}$ were plotted in temperature-magnetic field diagram (Fig. 6 (c), (d)). We estimated $H_{c2}(0)$ by the linear extrapolation of $T_c^{onset}$. The $H_{c2}(0)$ was surprisingly increased from 37 T to 72 T by applying a pressure of 1.48 GPa. The pressure effect for the enhancement of $H_{c2}$ is also remarkable in FeSe. Assuming that this superconductivity is in the dirty limit, $H_{c2}(0)$ is estimated to be ~ 50 T. Due to the high upper critical filed, this system is a candidate for materials of superconducting wire applications. More studies on FeSe system must be important in terms of basic physics and application of superconductivity.

In summary, we investigated the superconducting properties of a new superconductor FeSe and confirmed the huge pressure coefficient of 9.1 K/GPa. We also observed that the transition becomes sharp around 0.42 GPa. This will be a clue to finding the most ideal condition for superconductivity in iron-based systems. Estimated $H_{c2}$ under pressure is strongly enhanced compared to that at 0.00 GPa. The huge $H_{c2}$ indicates the possibility of application in materials for superconducting wires.

The authors would like to thank M. Nagao (Univ. of Yamanashi), A. Kikkawa, H. Suzuki and H. Kitazawa (NIMS) for useful discussions and experimental helps. This work was partly supported by Grant-in-Aid for scientific Research (KAKENHI).

References

[1] J. Nagamatsu, N. Nakagawa, T. Muranaka, Y. Zenitani, and J. Akimitsu, Nature **410**, 63 (2001).

[2] Y. Takano, H. Takeya, H. Fujii, H. Kumakura, T. Hatano, and K. Togano, Appl. Phys. Lett. **78**, 2914 (2001).

[3] Y. Takano, M. Nagao, I. Sakaguchi, M. Tachiki, T. Hatano, K. Kobayashi, H. Umezawa, H. Kawarada, Appl. Phys. Lett. **85**, 2851 (2004).

[4] Y. Kamihara, T. Watanabe, M. Hirano and H. Hosono, J. Am. Chem. Soc. **130**, 3296 (2008).

[5] H. Takahashi, K. Igawa, K. Arii, Y. Kamihara, M. Hirano and H, Hosono, nature **453**, 376 (2008).

[6] Z. A. Ren, J. Yang, W. Liu, W. Yi, X. L. Shen, Z. C. Li, G. C. Che, X. L. Dong, L. L. Sun, F. Zhou and Z. X. Zhao, Europhys. Lett. **82**, 57002 (2008).

[7] X. H. Chen, T. Wu, G. Wu, R. H. Liu, H. Chen and D. F. Fang, nature **453**, 761 (2008).

[8] M. Rotter, M. Tegel and D. Johrendt, arXiv:0805.4630 (2008).

[9] X. C. Wang, Q. Q. Liu, Y. X. Lv, W. B. Gao, L. X. Yang, R. C. Yu, F. Y. Li and C. Q. Jin, arXiv:0806.4688 (2008).

[10] I. A. Nekrasov, Z. V. Pchelkina and M. V. Sadovskii, arXiv:0807.2228 (2008).

[11] F. C. Hsu, J. Y. Luo, K. W. The, T. K. Chen, T. W. Huang, P. M. Wu, Y. C. Lee, Y. L. Huang, Y. Y. Chu, D. C. Yan and M. K. Wu, arXiv:0807.2369 (2008).

[12] G. Hägg and A. L. Kindstrom, Z. Phys. Chem. **22**, 455 (1933).

[13] D. A. Zocco, J. J. Hamlin, R. E. Baumbach, M. B. Maple, M. A. McGuire, A. S. Sefat, B. C. Sales, R. Jin, D. Mandrus, J. R. Jeffries, S. T. Weir and Y. K. Vohra, arXiv:0805.4372 (2008).

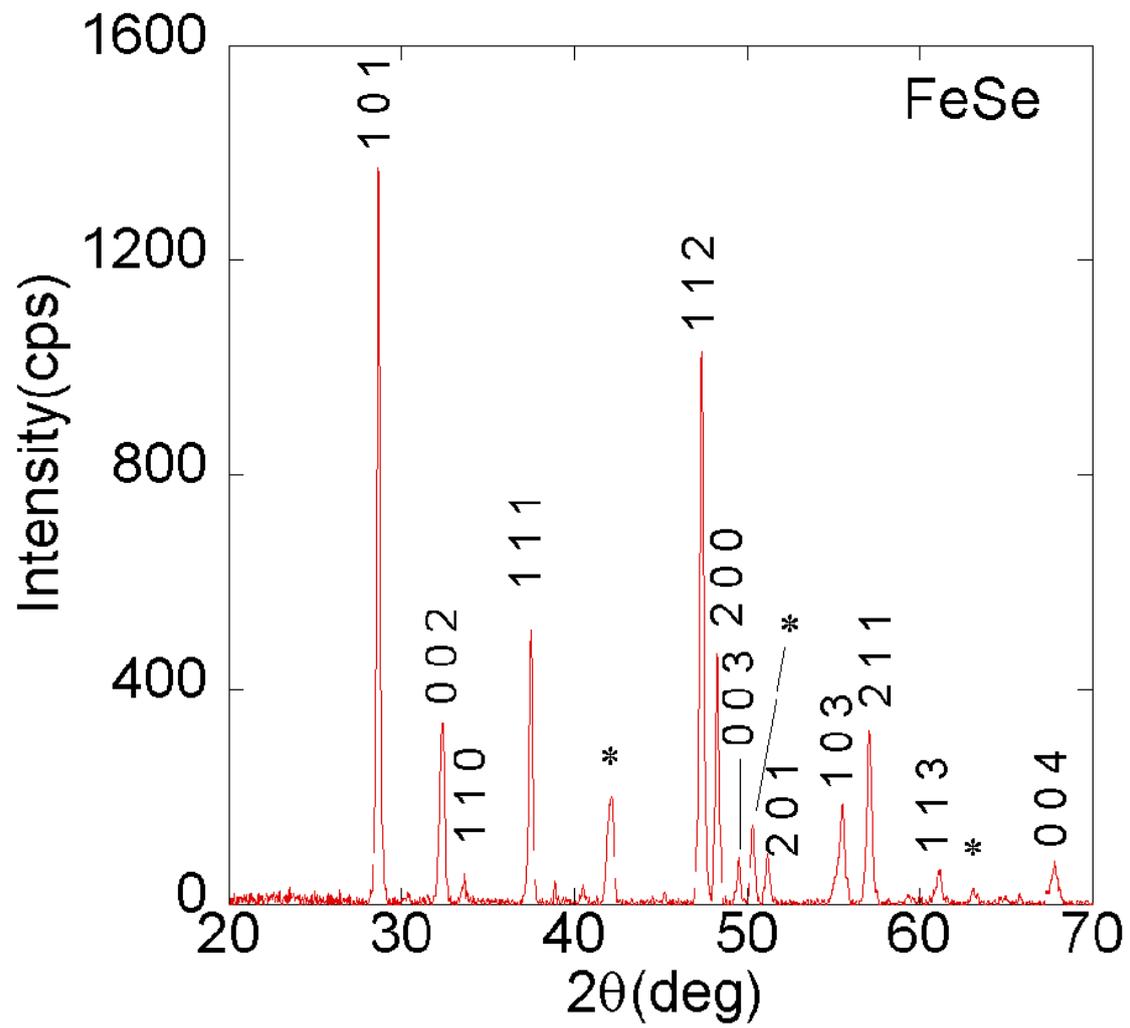

Fig. 1. X-ray diffraction pattern of FeSe with plane indices. The asterisks indicate the peaks of hexagonal phase.

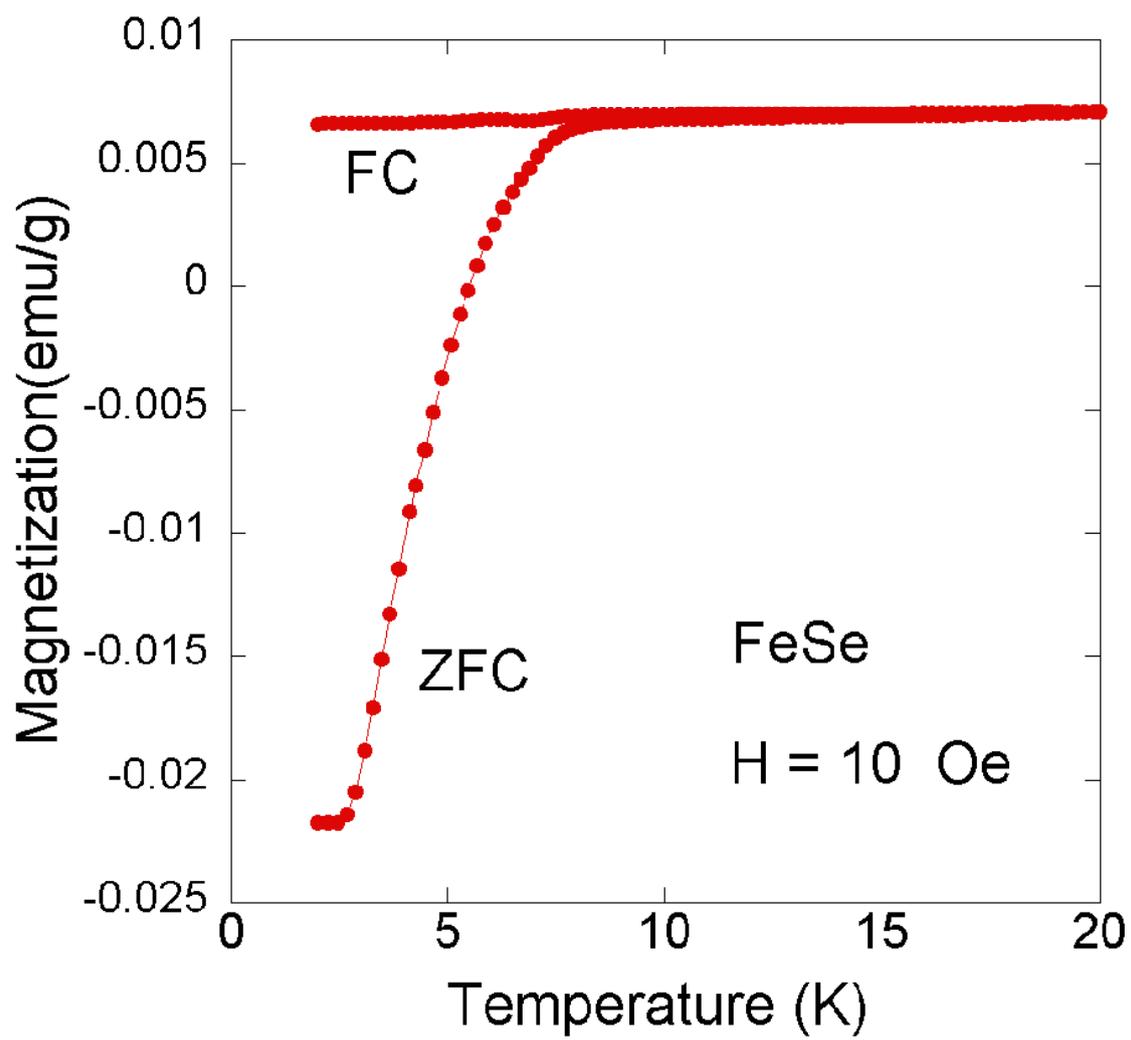

Fig. 2. Temperature dependence of magnetization.

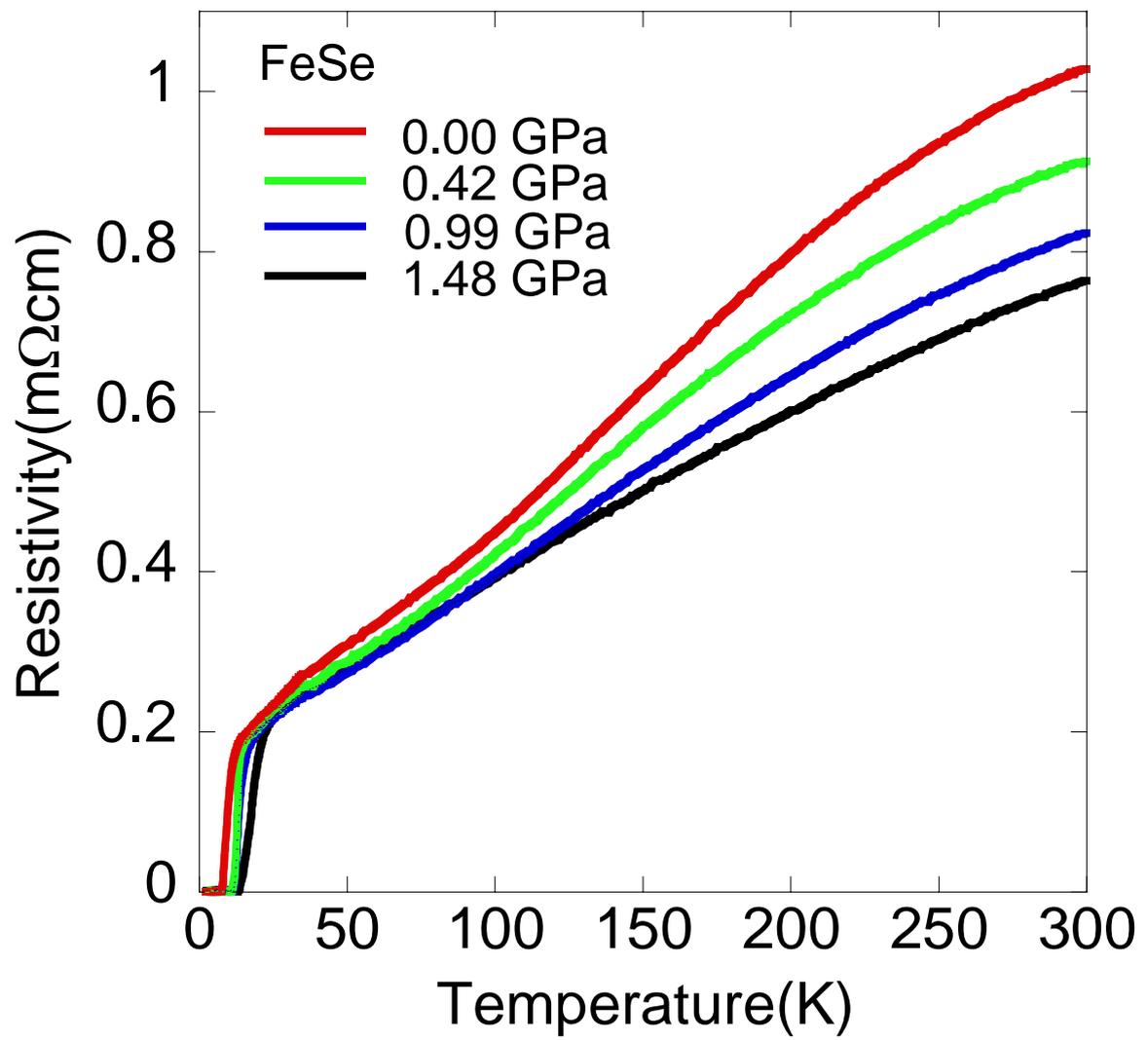

Fig. 3. Temperature dependence of resistivity under pressure.

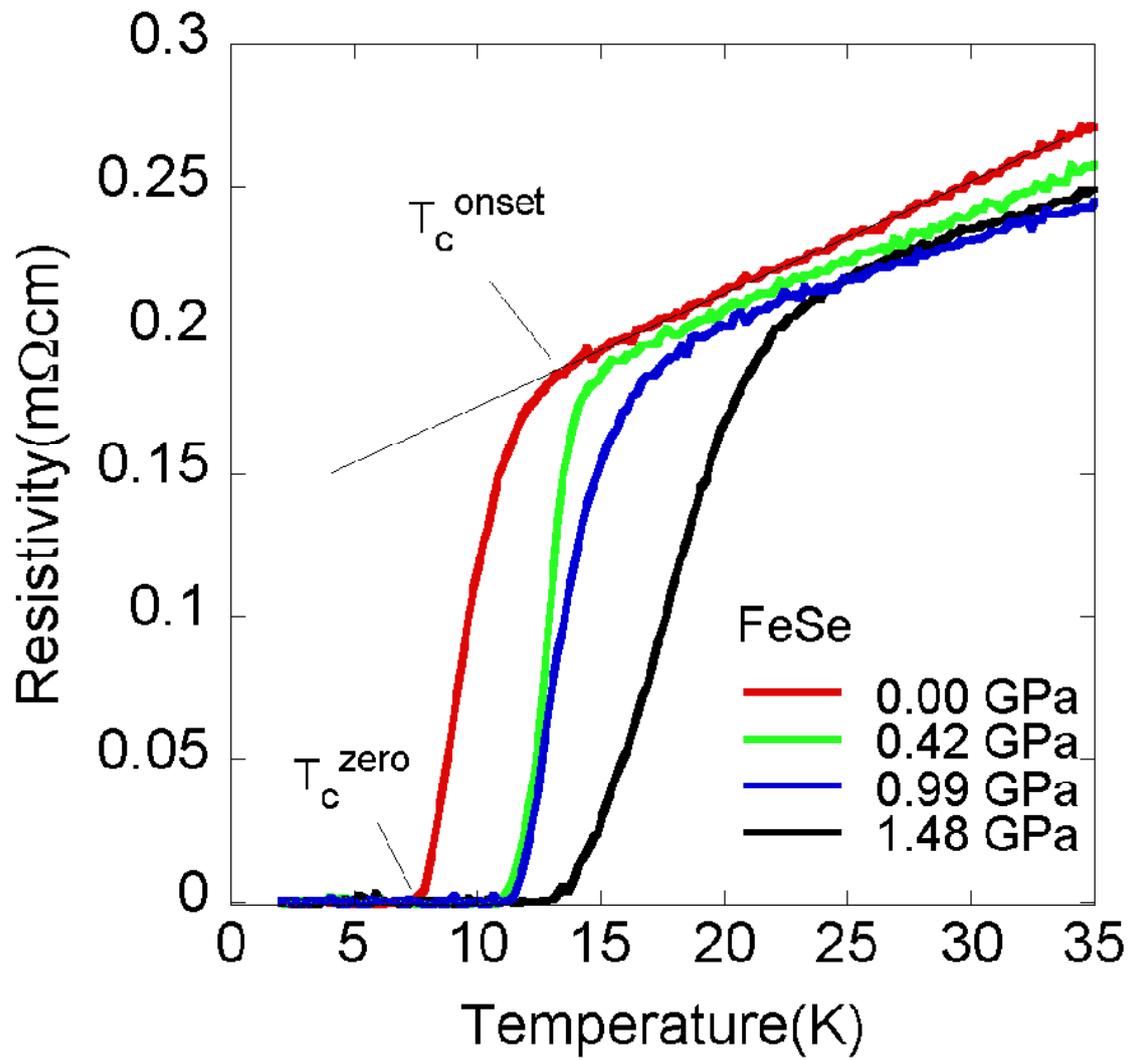

Fig. 4. Temperature dependence of resistivity around superconducting transition temperature under pressure.

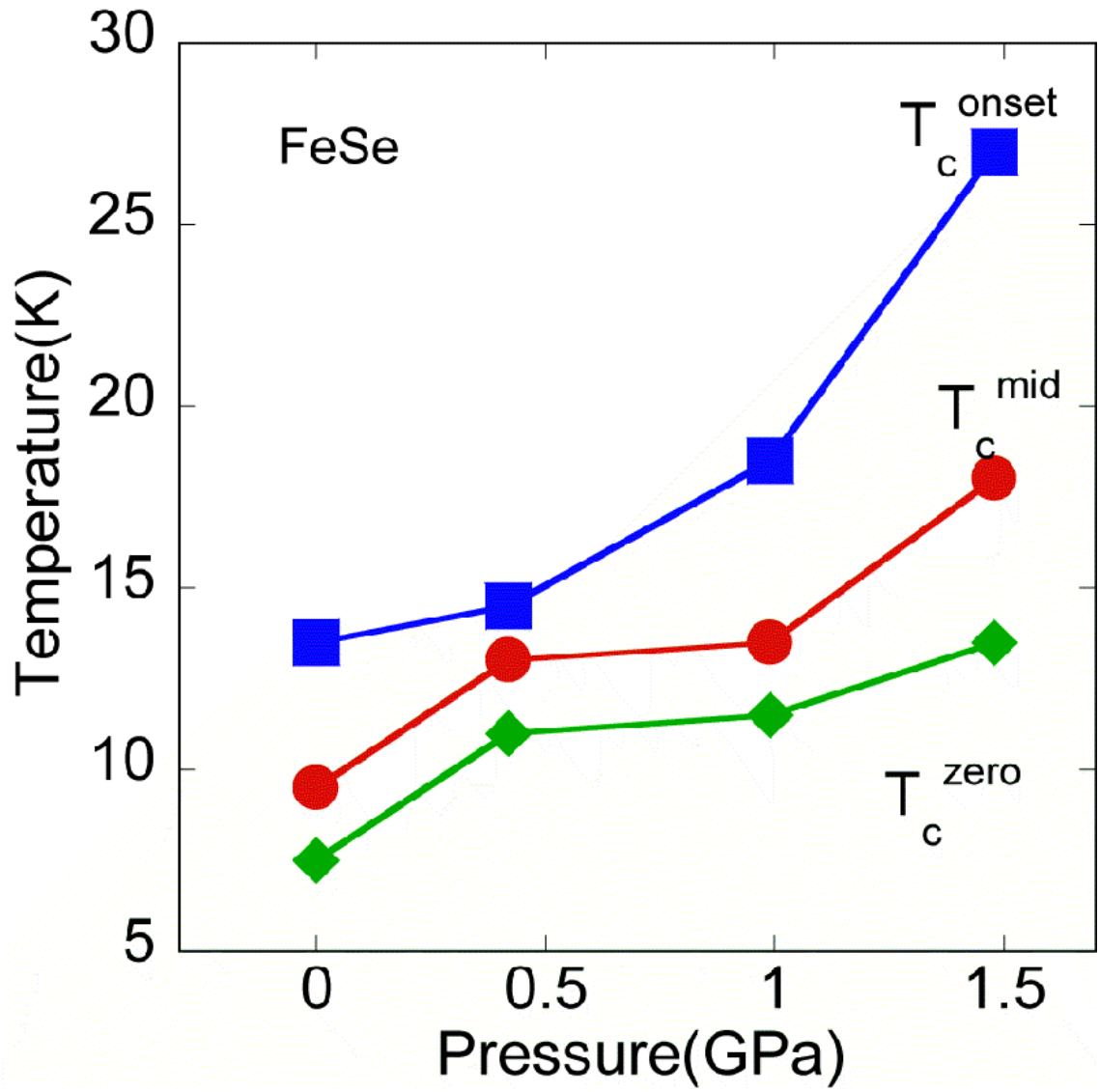

Fig. 5. Pressure dependence of $T_c^{onset}$, $T_c^{mid}$ and $T_c^{zero}$.

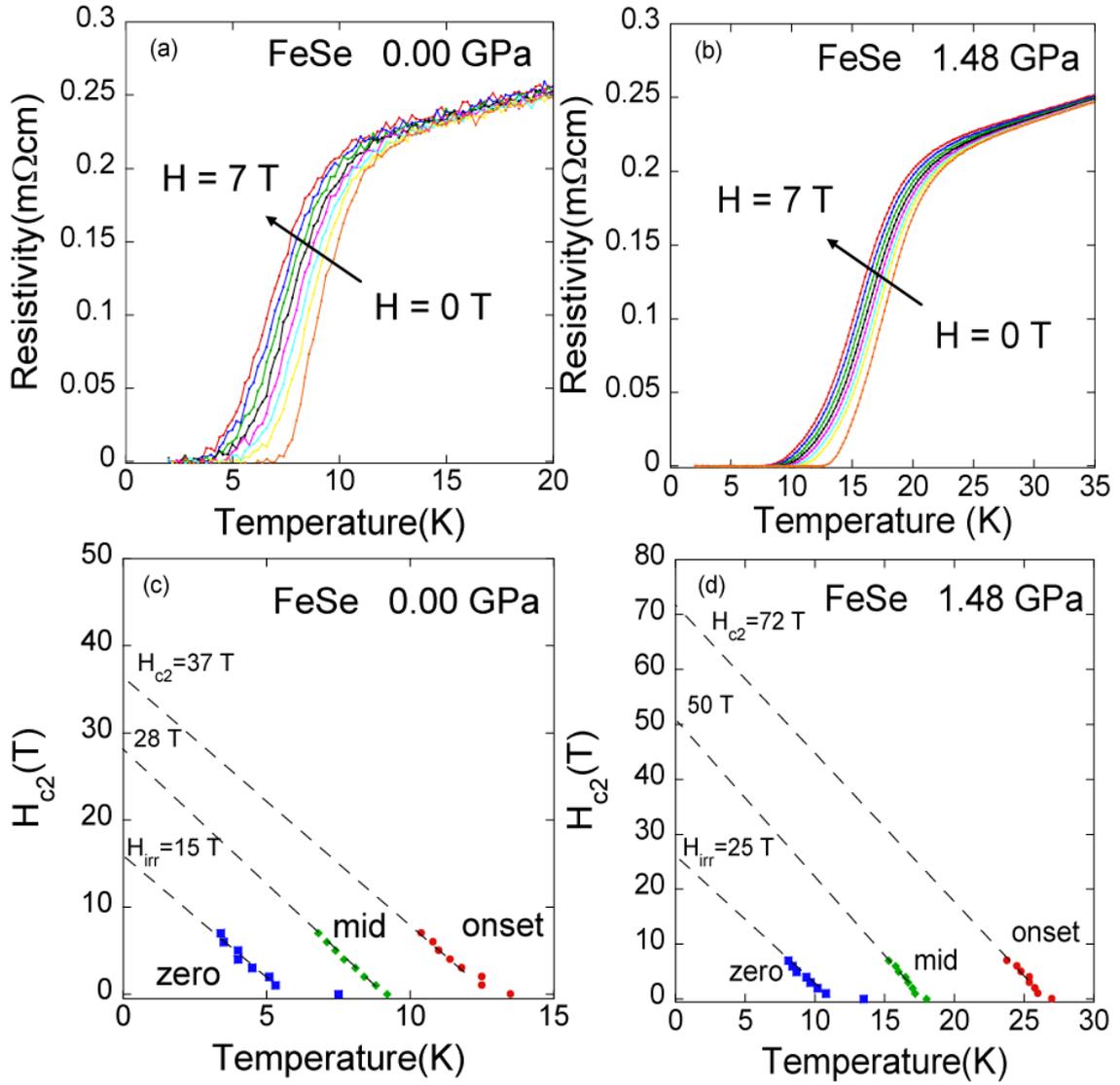

Fig. 6. (a)(b) Temperature dependence of resistivity at 0.00 GPa and 1.48 GPa under magnetic fields up to 7 T with an increment of 1 T. (c)(d) $H_{c2}(T)$ and $H_{irr}(T)$ plot defined by $T_c^{onset}$, $T_c^{mid}$ and $T_c^{zero}$ at 0.00 GPa and 1.48 GPa.